\documentclass[article,onecolumn,superscriptaddress,nofootinbib]{revtex4}
\usepackage{xcolor,color}
\usepackage[a4paper,
            bindingoffset=0.2in,
            left=.7in,
            right=1in,
            top=1in,
            bottom=1in,
            footskip=.25in]{geometry}
\usepackage[utf8]{inputenc}
\usepackage{float}
\usepackage[T1]{fontenc}
\usepackage{latexsym,amssymb,amsmath,amsfonts}
\usepackage{subfigure}
\usepackage{graphicx}
\usepackage{mathrsfs}
\usepackage{float}
\usepackage{hyperref}
\hypersetup{colorlinks=true, linkcolor=magenta, citecolor=green }
\usepackage{enumerate}

\begin{document}
\title{Selecting energy-momentum trace dependent gravity theories with LSS}
\author{Jonas Pinheiro da Silva\href{https://orcid.org/0000-0001-8456-0096}{\includegraphics[width=10pt]{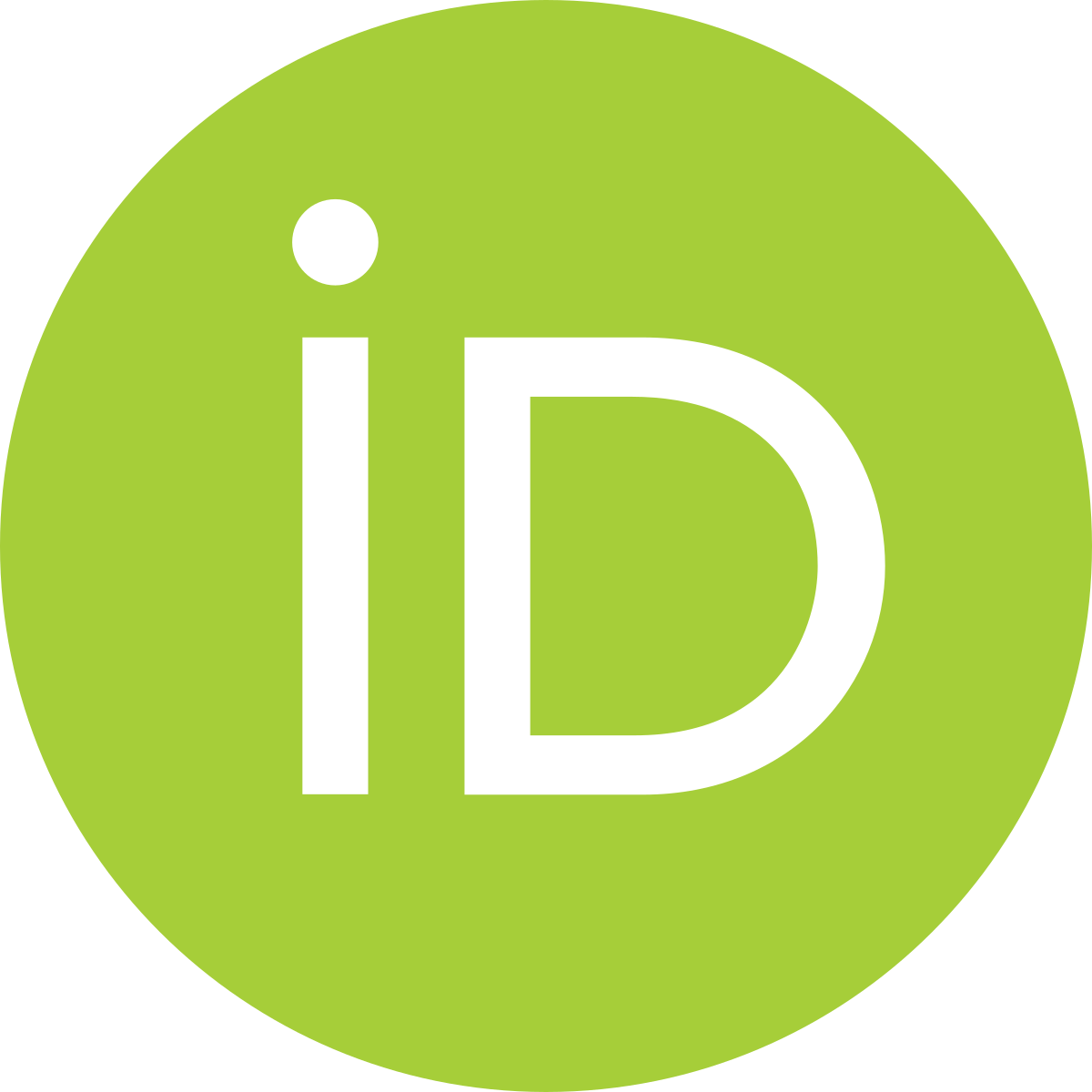}}}
\email{{jonas.j.silva@edu.ufes.br}}
\affiliation{PPGCosmo, CCE, Universidade Federal do Espírito Santo (UFES)\\ Av. Fernando Ferrari, 540, CEP 29.075-910, Vitória, ES, Brasil.}
\author{Hermano Velten\href{https://orcid.org/0000-0002-5155-7998 }{\includegraphics[width=10pt]{orcidID.png}}}
\email{hermano.velten@ufop.edu.br}
\affiliation{Departamento de F\'isica, Universidade Federal de Ouro Preto (UFOP), Campus Universit\'ario Morro do Cruzeiro, 35.402-136, Ouro Preto, Brazil}
\date{\today}
\begin{abstract}We study scalar cosmological perturbations in $f(R, T)$ modified gravity theories, where $T$ represents the trace of the energy-momentum tensor. We provide detailed equations for the matter energy density contrast. We solve them numerically to facilitate a comparison with available large-scale structure (LSS) formation observational data on $f \sigma_8$, while also addressing the $S_8$ tension. We identify $f(R,T)$ models that either lead to growth enhancement or suppression. Since recent results in the literature indicate a preference for the latter feature, this type of analysis is quite useful for selecting viable modifications of gravity. The studied class of such $f(R,T)$ models are either ruled out or severely restricted. After selecting the surviving $f(R, T)$ models we show how they deal with the $S_8$ tension.
\end{abstract}

\maketitle

\section{Introduction}
A number of considerable efforts have been devoted to assessing the validity of $f(R,T)$ theories, where ($T\equiv g^{\mu\nu}T_{\mu\nu}$) means the trace of the energy-momentum tensor, with special emphasis on the late time cosmological evolution. Since in such approach there is no associated component to the dark energy fluid, the pressureless matter,  presumably a Cold Dark Matter (CDM) component, influenced by this modified gravity theory is responsible for driving the accelerated expansion of the universe \cite{Velten:2017hhf, Moraes:2019hgx,  Jeakel:2023xlp, Myrzakulov:2023lcl, Iosifidis:2021kqo,Fortunato:2023ypc, Mishra:2024uwq, Bhattacharjee:2019oim, Bouali:2023fid}. However, it is essential to recognize that recent research has imposed substantial limitations on the practical usefulness of $f(R,T)$ approaches at the solar system level \cite{Bertini:2023pmp}. On the other hand, in the cosmological scenario, the results highlighted in Ref. \cite{Jeakel:2023xlp} are of particular interest since they demonstrate that though $f(R,T)$ theories can be consistent on late cosmological times, when a radiation component is added to the total cosmic energy budget the ability of these theories to maintain coherence fails.

The formulation of $f(R,T)$ theories from its Lagrangian to the field equations level is still a matter of debate. Even the standard practise to move the extra modified gravity terms from the left hand side to the right hand side of the field equations forming an effective energy-momentum tensor is disputed \cite{Fisher:2019ekh, PhysRevD.101.108501, Fisher:2020zwx}. Also, issues about the definition of the coupling term between the trace of the energy-momentum tensor and geometry are intrinsic to the formulation of $f(R,T)$ theories. In particular, there is the claim that trace free components do not couple to geometry as discussed in Ref. \cite{Shabani:2014xvi}. The present work is not dedicated to an exhaustive analysis of this aspect. Instead, we direct our focus to the formulation of the dynamical equations that dictate the behavior of cosmological matter scalar perturbations within the context of $f(R,T)$ theories. Therefore, in our approach, by neglecting the influence of radiation the issue about the coupling term is irrelevant. In fact, 
concerning only the growth of matter density fluctuations in an expanding universe, the cosmological matter dominated phase plays a crucial role in the final large scale structure patterns. According to the standard cosmological model, baryonic perturbations start to follow the already evolved CDM fluctuations after the last scattering surface when the background dynamics is dominated by a Einstein-de Sitter (matter like) expansion. Therefore, in order to analyse the matter (dark plus baryonic) clustering features in the observed LSS the radiative contribution is irrelevant for our purposes.

We expand previous analysis performed in Ref.\cite{Alvarenga:2013syu} by considering now the non-conservative aspect of the continuity equation, an issue avoided in such reference. We also obtain first order dynamical equations for the scalar perturbations which are slightly different from those obtained in \cite{Alvarenga:2013syu}. The relevance of this formulation emerges as it offers a more complete and precise exploration of the observable manifestations that arise during the process of structure formation in a cosmological universe dominated by matter under the influence of energy-momentum trace dependent modified gravity theories. Given the growth of the density contrast, an inherent feature of a matter-dominated universe, our investigation turns to a deep exploration of the free model parameter region where the $f(R,T)$ models remain theoretically sound. Therefore, the main goal of this study lies in producing results that shed light on the status of these theories concerning the structure formation process. 

In the next section \ref{s2} we provide a brief analytical development for obtaining the field equations in the $f(R, T)$ family of theories. We also present the first order expressions for the field equations which shall be used later. Subsequently, in section \ref{s3} we revisit the cosmological scalar perturbations equations in $f(R, T)$ theories. In particular, we obtain solutions for a perfect fluid cosmological component. We then solve such equations numerically and promote a comparison with currant available observational data. This analysis will be very important to select viable $f(R,T)$ models. The final section is devoted to our conclusions.

\section{General framework}\label{s2}
The general action for the $f(R, T)$ theories is written according to the general action \cite{Harko:2011kv}, 
\begin{equation}\label{action}
    S\left(g^{\mu\nu}R_{\mu\nu}, g^{\mu\nu}T_{\mu\nu}, \psi _{m}\right) = \int d ^{4}x\sqrt{-g}\left[L_{G}\left(R, T\right) + L_{m}(g_{\mu\nu}, \psi _{m})\right],
\end{equation}
where the Lagrangian density for the gravitational sector, $L_{G} = f(R, T)/\kappa ^{2}$, depends on the curvature scalar ($R\equiv g^{\mu\nu}R_{\mu\nu}$) and on the energy-momentum tensor trace $T$. The coupling constant is defined exactly as in General Relativity (GR) i.e., $\kappa ^{2} = 8\pi G$. All matter fields are denoted by $\psi_m$. In turn, the energy-momentum tensor is obtained as usual 
\begin{equation}\label{ttotal}
    T _{\mu\nu} \equiv - \frac{2}{\sqrt{-g}}\frac{\delta \left(L_{m}\sqrt{-g}\right)}{\delta g^{\mu\nu}}.
\end{equation}

 The field equations are obtained after variation of action \eqref{action}. In this process, one has to obtain the expression for $\delta f(R,T)$. In order to focus on the contribution of the energy-momentum tensor let us now follow the arguments presented in Ref. \cite{Shabani:2014xvi}.  The total energy-momentum tensor \eqref{ttotal} can be split into relativistic ($L^{(r)}$) and non-relativistic ($L^{(m)}$) components
\begin{equation}
    L_{m} \rightarrow L^{(m)} + L^{(r)},
\end{equation}
then, Eq. \eqref{ttotal} is rewritten as
\begin{equation}\label{tem1}
    T_{\mu\nu} = T^{(m)}_{\mu\nu} + T^{(r)}_{\mu\nu} = - \frac{2}{\sqrt{-g}}\frac{\delta \left(\sqrt{-g}L^{(m)}\right)}{\delta g^{\mu\nu}} - \frac{2}{\sqrt{-g}}\frac{\delta \left(\sqrt{-g}L^{(r)}\right)}{\delta g^{\mu\nu}}.
\end{equation}
After varying the action \eqref{action} with respect to the metric, the contribution from the energy-momentum tensor is encoded in the term, 
\begin{eqnarray}\label{dt0}
    \delta f(T) = f_{T}\delta T = f_{T}\delta \left(g^{\mu\nu}T_{\mu\nu}\right).
\end{eqnarray}
Since $g^{\mu\nu}T^{(r)}_{\mu\nu} = 0$, then
\begin{equation}
    \frac{\delta f(T)}{\delta g^{\mu\nu}} = f_{T}\left(T^{(m)}_{\mu\nu} + g^{\mu\nu}\frac{\delta T^{(m)}_{\mu\nu}}{\delta g^{\mu\nu}}\right).
\end{equation}

This means that only the non-relativistic matter couples to the geometric sector in $f(R,T)$ gravity. This is a quite important issue in this sort of theory. 

Before we proceed, a cautionary remark is necessary. In a previous work by the same authors \cite{Jeakel:2023xlp}, we analysed the FLRW late-time background expansion of $f(R,T)$ cosmologies. Regarding the late-time phenomenology, we found that $f(R,T)$ cosmologies have a narrow free model parameter space in agreement with current observations. However, in such reference, when we attempted to analyse the complete cosmological scenario, including the early-time aspects, the inclusion of radiation has been made by adding the relativistic pressure at the field equations level i.e., we replaced the total pressure $p$ by the sum of non-relativistic and relativistic components $p_m+p_r$. Consequently, we found the inviability of $f(R,T)$ cosmologies when considering the radiation fluid, as this component would no longer scale according to $\rho_r \propto a^{-4}$ as usual, due to the non-conservation of the energy-momentum tensor in $f(R,T)$. In summary, the conclusions presented in \cite{Jeakel:2023xlp} are irrelevant for the purposes of the present work, as we will not take into account relativistic components into account here.

The action \eqref{action} leads to the following field equations\footnote{The coupling between matter and curvature implies in the  existence of a matter tensor $\Theta _{\mu\nu}$ (see \cite{Harko:2011kv} for details). In this case, we write $\Theta _{\mu\nu}^{m} = g^{\mu\nu}\frac{\delta T^{m}_{\mu\nu}}{\delta g^{\mu\nu}}$. Due to the perfect fluid structure, one  has two possibilities for the matter Lagrangian: $L_{m} = \rho$ or $L_{m} = -p$. A detailed discussion about the choice of the Lagrangian and its implications on the dynamics is found in Refs.\cite{Sotiriou:2008it, Bertolami:2008ab, Faraoni:2009rk}.}
\begin{equation}\label{eq1}
    f_{R}R^{\mu}_{\nu} - \frac{f(R, T)}{2}\delta ^{\mu}_{\nu} - \Delta ^{\mu}_{\nu}f_{R} = \kappa ^{2}T^{\mu}_{\nu} + f_{T}T^{\mu (m)}_{\nu}. 
\end{equation}
The term $\Delta ^{\mu}_{\nu}$, is defined as a combination of operators
\begin{equation}
    \Delta ^{\mu}_{\nu} \equiv g ^{\mu\alpha}\nabla _{\alpha}\nabla _{\nu} - \delta ^{\mu}_{\nu}g ^{\alpha \beta}\nabla _{\alpha}\nabla _{\beta}. 
\end{equation}
The terms proportional to $f_{T}$ in equation \eqref{eq1} are the geometry-matter coupling terms that arises from the variation of the trace of the energy-momentum tensor. It is worth noting that for a pressureless fluid ($p_{m}=0$), mimicking a matter dominated universe, the right hand side of \eqref{eq1} is written simple as $(\kappa^2+f_T)T^{\mu (m)}_{\nu}$ i.e., the combination $(\kappa^2+f_T)$ becomes the effective coupling between the energy-momentum tensor and geometry. 

A complete description of $f(R, T)$ theories in the metric and metric-affine formalism can be found, respectively, in Refs.\cite{Harko:2011kv, Barrientos:2018cnx}. 

An important feature of the $f(R, T)$ class of  theories is its non-conservative nature. Generally, the presence of matter and (or) coupling with second order terms in the $R$-curvature in the Lagrangian density of the gravitational sector implies in such non-conservation. Via the Bianchi identity $\nabla _{\mu}G^{\mu}_{\nu} = 0$ (the restricted case where $f(R,T)$ depends linearly on $R$ plus a trace dependence) and propriety's from Levi-Civita connection, $\nabla _{\mu}g^{\mu\nu} = 0$, one finds that equation \eqref{eq1} turns into
\begin{equation}\label{eq1.1}
    \nabla _{\mu}T^{\mu (m)}_{\nu} = - \frac{1}{(\kappa ^{2} + f_{T})}\left(\nabla _{\mu}f_{T}T^{\mu (m)}_{\nu} + f_{T}\frac{1}{2}\nabla _{\mu}T\right).
\end{equation}

Once more, conservative $f(R,T)$ models can be conceived by forcing the right hand side of the above expression to vanish. For pressureless fluids this happens when $f(R,T)=R + T^{1/2}$. This is the case studied in Ref. \cite{Alvarenga:2013syu}.

Hereafter, we will avoid using the index $m$ for non-relativistic matter. Then, all equations below apply to a universe fully dominated by non-relativistic matter.

\subsection{Perturbed field equations}

Since we will be interested in the perturbative dynamics of $f(R,T)$ theories, let us present the first order perturbed version of the field equation \eqref{eq1}.  It reads 
\begin{eqnarray}\label{eq3}
    \delta G^{\mu}_{\nu} = \kappa ^{2}\delta T^{\mu}_{\nu} + 
  f_{\bar{T}\bar{T}}\delta T\left(\bar{T}^{\mu}_{\nu} + \bar{p}\delta ^{\mu}_{\nu}\right) + f_{\bar{T}}\left(\delta T^{\mu}_{\nu}+ \delta p \delta ^{\mu}_{\nu} + \frac{1}{2}\delta T\delta ^{\mu}_{\nu}\right).
\end{eqnarray}
From now on all non-perturbed (background) terms carry an overbar.
In order to arrive to the above equation we have assumed the minimally coupled scenario
$f(R, T) = R + f(T)$.  In this case $f_{R} = \partial f(R,T)/\partial R = 1$. Correspondingly, the derivative $f_{T}=\partial f(R,T)/\partial T$ is written at first order as $\delta f_{T} = f_{\bar{T}\bar{T}}\delta T$.

From \eqref{eq1.1}, the general form of the first order continuity equation is written as
\begin{eqnarray}\label{eq6}
    &&\left(\partial _{\mu}\delta T^{\mu}_{\nu} + \delta \Gamma ^{\mu}_{\mu\lambda}\bar{T}^{\lambda}_{\nu} + \bar{\Gamma}^{\mu}_{\mu\lambda}\delta T^{\lambda}_{\nu} - \delta \Gamma ^{\lambda}_{\mu\nu}\bar{T}^{\mu}_{\lambda} - \bar{\Gamma} ^{\lambda}_{\mu\nu}\delta T^{\mu} _{\lambda}\right)(\kappa ^{2} + f_{\bar{T}})  = - \biggl[ f_{\bar{T}\bar{T}} \nabla _{\mu}\bar{T}^{\mu}_{\nu} \delta T + f_{\bar{T}\bar{T}}\partial _{\mu}\bar{T}\left(\delta T^{\mu}_{\nu} + \delta p \delta ^{\mu}_{\nu}\right) \nonumber \\
    && +\left(f_{\bar{T}\bar{T}\bar{T}}\partial _{\mu}\bar{T}\delta T + f_{\bar{T}\bar{T}}\partial _{\mu}\delta T\right)\left(\bar{T}^{\mu}_{\nu} + \bar{p}\delta ^{\mu}_{\nu}\right) + f_{\bar{T}\bar{T}}\delta T\left(\frac{1}{2}\partial _{\mu}\bar{T} + \partial _{\mu}\bar{p}\delta ^{\mu}_{\nu}\right) + f_{\bar{T}}\left(\frac{1}{2}\partial _{\mu}\delta T + \partial _{\mu}\delta p \delta ^{\mu}_{\nu}\right)\biggr].
\end{eqnarray}
This will be used in the next section.

\section{Scalar perturbations for a perfect fluid}\label{s3}

Cosmological perturbation theory is designed to indicate spacetime anisotropies and inhomogeneities that are not considered in the Friedmann-Lemaître-Robertson-Walker (FLRW) metric. Thus, we can define the FLRW (flat) background metric, $\bar{g}_{\mu\nu}$, as 
\begin{equation}\label{p0}
    \Bar{g}_{\mu\nu} \equiv g_{\mu\nu} - \delta g_{\mu\nu},
\end{equation}
where in this case, $g_{\mu\nu}$ denotes the metric of the FLRW-perturbed spacetime with the contribution of the perturbations $\delta g_{\mu\nu}$. In terms of the conformal time $\tau$, the FLRW metric reads
\begin{equation}\label{mc}
    d\bar{s}^{2} = a(\tau) ^{2}\left(d\tau ^{2}-\delta _{ij}dx^{i}dx^{j}\right),
\end{equation}
Via \eqref{p0} this leads us to the FLRW-perturbed metric 
\begin{equation}\label{p4}
    ds^{2} = a(\tau) ^{2}\eta _{\mu\nu} + \delta g_{\mu\nu},
\end{equation}
where $\eta _{\mu\nu}$ is the Minkowski metric. As we are only interested in the first order scalar perturbation structure, let us fix the gauge
\begin{equation}\label{p3}
\delta g_{\mu\nu} = 2a(\tau)^{2}
\left(\begin{array}{cc}
\phi &0\\
0 & \psi \delta _{ij} \\
\end{array}\right),
\end{equation}
hence, the line element \eqref{p4} becomes
\begin{equation}\label{m3}
    ds^{2} = a(\tau)^{2}\left[(1 + 2\phi)d\tau ^{2} - ( 1 - 2\psi )\delta _{ij}dx^{i}dx^{j}\right].
\end{equation}
With the metric signature used here we have the same line element used as used e.g., in \cite{Mukhanov:1988jd, Alvarenga:2013syu}. In literature the line element \eqref{m3} is called of conformal Newtonian gauge or longitudinal gauge. The components of perturbed energy-momentum tensor for cosmological perfect fluids in this gauge is given by 
\begin{equation}
    \delta T ^{0}_{0} = \delta \rho; \hspace{0.5cm} \delta T^{i}_{j} = -\delta p \delta ^{i}_{j}; \hspace{0.5cm} \text{and} \hspace{0.5cm} \delta T^{0}_{i} = -\delta T^{i}_{0} = - (1 + c_{s}^{2})\bar{\rho}v_{i},
\end{equation}
being $\bar{\rho}$ and $\bar{p}$ the background contribution of the fluid's energy density and pressure, respectively, connected by the equation of state parameter $\bar{\omega}=\bar{p}/\bar{\rho}$. The four velocity is defined as $v_{\mu}=(0, v_i)$. Developing the right side of the \eqref{eq3} one finds
\begin{eqnarray}\label{eq4}
     \delta G^{\mu}_{\nu}  = \left(\kappa ^{2} + f_{\bar{T}}\right)\delta T^{\mu}_{\nu} +  \left[f_{\bar{T}\bar{T}}(1 - 3c^{2}_{s})(1 + \bar{\omega})\bar{\rho}v^{\mu}v_{\nu} + \frac{f_{\bar{T}}}{2}(1 - c^{2}_{s})\delta ^{\mu}_{\nu}\right]\delta \rho. 
\end{eqnarray}
For the sake of generality, let us consider an equation of state $\bar{\omega}\equiv \bar{\omega} (\bar{\rho})$. Therefore, an adiabatic process is defined by the relation 
\begin{equation}
    \delta p = \left(\bar{\omega} + \bar{\rho} \frac{d\bar{\omega}}{d\bar{\rho}}\right)\delta \rho = c^{2}_{s}\delta \rho,
    \end{equation}
where $c^2_{s}$ is the square sound speed. If one adopts a constant equation of state parameter $\bar{\omega}$, then $\bar{\omega}=c^2_s$.
After such definitions and trivial calculations, the $\{00\}$, $\{0i\}$ and $\{ij\}$ components of \eqref{eq4} are, respectively
\begin{equation}\label{poisson}
    \nabla ^{2}\psi - 3\mathcal{H}\left(\psi ^{\prime} + \mathcal{H}\phi\right) = \frac{a^{2}(\eta)}{2}\kappa ^{2}\bar{\rho}\left\{1 + \frac{1}{\kappa ^{2}}\left[f_{\bar{T}\bar{T}}(1 - 3c^{2}_{s})(1 + c^{2}_{s})\bar{\rho} + \frac{f_{\bar{T}}}{2}(3 - c^{2}_{s})\right]\right\}\delta,
\end{equation}
\begin{equation}\label{ve}
(\phi ^{\prime} + \mathcal{H}\phi)_{,i} = \frac{a^{2}(\eta)}{2}\kappa ^{2}\bar{\rho}(1 + \bar{\omega})\left(1 + \frac{f_{\bar{T}}}{\kappa ^{2}}\right)v_{i},
\end{equation}
\begin{equation}\label{potential}
\phi ^{\prime\prime} + 3\mathcal{H}\phi ^{\prime} + (2\mathcal{H}^{\prime} + \mathcal{H}^{2})\phi = \frac{a^{2}(\eta)}{2}\kappa ^{2}\bar{\rho}\left[c^{2}_{s} - \frac{f_{\bar{T}}}{2\kappa ^{2}}(1 - 3c^{2}_{s})\right]\delta, \hspace{0.5cm} \text{(if $i = j$)},
\end{equation}
and
\begin{equation}\label{eq5}
    \left(\phi - \psi \right)_{, ij} = 0 \hspace{0.5cm} \text{(if $i \neq j$)};
\end{equation}
being the matter density contrast defined as $\delta\equiv \delta\rho/ \bar{\rho}$. Note that for perfect (shear free) fluids we only have one degree of freedom associated with the scalar perturbed metric potentials (the result $\phi=\psi$ in \eqref{eq5} arises naturally).  

Elaborating further the continuity equation \eqref{eq6} we find for the $\nabla _{\mu}T_{0}^{\mu}$ component, in Fourier space ($\nabla^2 \rightarrow -k^2$, being $k$ the wavenumber),
\begin{eqnarray}\label{eq7.1}
&&(\delta \rho)^{\prime} + 3\mathcal{H}(1 + c_{s}^{2})\delta \rho - (1 + \bar{\omega})(k ^{2}v + 3\phi ^{\prime})\bar{\rho} = \frac{(1 - 3c_{s}^{2})^{2}\bar{\rho}^{\prime}\bar{\rho}\delta}{(\kappa ^{2} + f_{\bar{T}})^{2}}\left[(1+\bar{\omega})f_{\bar{T}\bar{T}}\bar{\rho} + \frac{f_{\bar{T}}}{2} + \frac{f_{\bar{T}}c_{s}^{2}}{1 - 3c_{s}^{2}}\right] \nonumber \\
&&- \frac{(1 - 3c_{s}^{2})\bar{\rho}\delta ^{\prime}}{\kappa ^{2} + f_{\bar{T}}}\biggl[f_{\bar{T}\bar{T}}\bar{\rho} + \frac{1}{2}f_{\bar{T}} + \bar{\omega}f_{\bar{T}\bar{T}}\bar{\rho} + \frac{c_{s}^{2}f_{\bar{T}}}{1 - 3c_{s}^{2}}\biggr] - \frac{(1-3c_{s}^{2})^{2}\bar{\rho}^{\prime}\delta}{\kappa ^{2} + f_{\bar{T}}}\biggl\{f_{\bar{T}\bar{T}\bar{T}}\bar{\rho}^{2}(1 + \bar{\omega}) + \frac{1}{2}f_{\bar{T}\bar{T}}\bar{\rho} + \nonumber \\
&&\frac{1}{1 - 3c_{s}^{2}}\biggl[2f_{\bar{T}\bar{T}}\bar{\rho}(1 + \bar{\omega}) + \frac{1}{2}f_{\bar{T}} + c_{s}^{2}f_{\bar{T}\bar{T}}\bar{\rho} + \frac{c_{s}^{2}f_{\bar{T}}}{1 - 3c_{s}^{2}}\biggr]\biggr\}.
\end{eqnarray}
At the same time, the \textit{momentum} component $\nabla _{\mu}T^{\mu}_{i}$ reads, 
\begin{eqnarray}\label{eq7.2}
    && - (1 + \bar{\omega})\left[(\bar{\rho}v)^{\prime} + \bar{\rho}\phi + 4\bar{\rho}\mathcal{H}v\right] = - (1 + \bar{\omega})(1 - 3c_{s}^{2})f_{\bar{T}\bar{T}}\bar{\rho}\bar{\rho}^{\prime}v + \frac{(1 -3c_{s}^{2})}{2}\{[(1 - 3c_{s}^{2})f_{\bar{T}\bar{T}}\bar{\rho} + f_{\bar{T}}]\bar{\rho}^{\prime}\delta + f_{\bar{T}}\bar{\rho}\delta ^{\prime}\} + \nonumber \\
    && + \bar{\omega}(1 - 3c_{s}^{2})[f_{\bar{T}\bar{T}}\delta \rho \bar{\rho} ^{\prime} + f_{\bar{T}\bar{T}\bar{T}}\delta \rho \bar{\rho} ^{\prime} + f_{\bar{T}\bar{T}}\bar{\rho}(\delta \rho)^{\prime}].
\end{eqnarray}

Now, for the sake of simplicity, let us define the following combination 
\begin{equation}\label{csi}
    \bar{\xi} = 1 + \frac{1}{\kappa ^{2} + f_{\bar{T}}}\left(f_{\bar{T}\bar{T}}\bar{\rho} + \frac{1}{2}f_{\bar{T}}\right).
\end{equation}
By eliminating the $T$ dependence on $f(R,T)$ one gets $\bar{\xi}=1$. This happens in either GR or $f(R)$ theories.
This definition will appear below in several equations. The relevance of this definition is understood when analysing the conservation equation \eqref{eq1.1}. In fact, for the background FLRW metric and considering a pressureless fluid i.e., $\bar{w}= c_{s}^{2} = 0$, the continuity equation \eqref{eq1.1} is written as
\begin{equation}
    \bar{\xi} \bar{\rho}^{\prime}+3 \mathcal{H}\bar{\rho}=0.
    \label{conserxi}
\end{equation}
Again, $f(R,T) \propto T^{1/2}$ is the unique situation in which $\bar{\xi}=1$ and modified gravity contribution is switched on.

Back to the perturbed level, adopting a pressureless fluid and eliminating the $\bar{\rho}^{\prime}$ terms with help of \eqref{conserxi}, the Eqs. \eqref{eq7.1} and \eqref{eq7.2} turn into, respectively
\begin{equation}\label{eq7}
      \bar{\xi} \delta ^{\prime} - (k^{2}v + 3\phi ^{\prime}) = -\frac{3\mathcal{H}\bar{\rho}\delta}{\kappa ^{2} + f_{\bar{T}}}\left[f_{\bar{T}\bar{T}} - \frac{1}{\bar{\xi}}\left(f_{\bar{T}\bar{T}\bar{T}}\bar{\rho} + \frac{5}{2}f_{\bar{T}\bar{T}}\right)\right],
\end{equation}
and
\begin{equation}\label{eq7.3}
v^{\prime} + \phi + \mathcal{H}v\left[4 -\frac{3}{\bar{\xi}}\left(1 + \frac{f_{\bar{T}\bar{T}}\bar{\rho}}{\kappa ^{2} + f_{\bar{T}}}\right)\right] = \frac{1}{2(\kappa ^{2} + f_{\bar{T}})}\left[f_{\bar{T}}\delta ^{\prime} - \frac{3\mathcal{H}\delta}{\bar{\xi}}\left(f_{\bar{T}\bar{T}}\bar{\rho} + f_{\bar{T}}\right)\right].
\end{equation}

By combining the time derivative of Eq. \eqref{eq7} with \eqref{eq7.3} we obtain a general equation for the evolution of the matter energy density contrast $\delta$. It reads
\begin{eqnarray}\label{dc}
    &&\biggl\{\delta ^{\prime \prime} + \delta ^{\prime}\mathcal{H}\biggl[4-\frac{3}{\bar{\xi}}\biggl(1 + \frac{f_{\bar{T}\bar{T}}\bar{\rho}}{\kappa ^{2} + f_{\bar{T}}}\biggr)\biggr]\biggr\}\bar{\xi} + \frac{3}{\kappa ^{2} + f_{\bar{T}}}\biggl[f_{\bar{T}\bar{T}} - \frac{1}{\bar{\xi}}\left(f_{\bar{T}\bar{T}\bar{T}}\bar{\rho} + \frac{5}{2}f_{\bar{T}\bar{T}}\right)\biggr]\biggl[\mathcal{H}^{\prime} \bar{\rho}\delta + \delta ^{\prime}\mathcal{H}\bar{\rho} + \nonumber \\
    && + \frac{3\mathcal{H}^2\bar{\rho}}{\bar{\xi}} \delta\biggl(1 - \frac{f_{\bar{T}\bar{T}}}{\kappa ^{2} + f_{\bar{T}}}\biggr)\biggr]  +3\mathcal{H}\biggl[4-\frac{3}{\bar{\xi}}\biggl(1 + \frac{f_{\bar{T}\bar{T}}\bar{\rho}}{\kappa ^{2} + f_{\bar{T}}}\biggr)\biggr]\biggl\{\frac{\mathcal{H}\bar{\rho}\delta}{\kappa ^{2} + f_{T}}\biggl[f_{\bar{T}\bar{T}} - \frac{1}{\bar{\xi}}\left(f_{\bar{T}\bar{T}\bar{T}}\bar{\rho} + \frac{5}{2}f_{\bar{T}\bar{T}}\right)\biggr] - \phi ^{\prime}\biggr\} - \nonumber \\
    && -3\phi ^{\prime \prime} + \delta ^{\prime}\xi ^{\prime} + \frac{3\mathcal{H}\bar{\rho}\delta }{\kappa ^{2} + f_{\bar{T}}}\biggl[-\frac{3\mathcal{H}\bar{\rho}}{\bar{\xi}}f_{\bar{T}\bar{T}\bar{T}} +  \frac{\bar{\xi}^{\prime}}{\bar{\xi ^{2}}}\biggl(\frac{5}{2}f_{\bar{T}\bar{T}} + f_{\bar{T}\bar{T}\bar{T}}\bar{\rho}\biggr)+\frac{3\mathcal{H}\bar{\rho}}{\bar{\xi}^2}\biggl(f_{\bar{T}\bar{T}\bar{T}\bar{T}}\bar{\rho} + \frac{7}{2}f_{\bar{T}\bar{T}\bar{T}}\biggr)\biggr] + \nonumber \\
    &&+ k^{2}\biggl\{\phi - \frac{1}{2(\kappa ^{2} + f_{\bar{T}})}\biggl[f_{\bar{T}}\delta ^{\prime} - \frac{3\mathcal{H}\delta }{\bar{\xi}}(f_{\bar{T}\bar{T}}\bar{\rho}+f_{\bar{T}})\biggr]\biggr\} = 0.
\end{eqnarray}
General relativity is recovered by making $f(T) = 0$ and the $\Lambda$CDM limit is achieved with a specific choice of free model parameters.

Equation \eqref{dc} is the general equation for the matter energy density contrast in $f(R,T)$ models. In order to solve it and, since we will consider scales that are within the sound horizon, we apply to it the quasi-static approximation adopting that time derivatives of $\phi$ are subdominant with respect to spatial derivatives and we will also rewrite $\phi$ in \eqref{dc} in terms of the remaining scalar sector quantities using \eqref{poisson} and \eqref{ve} (see \cite{Maggiore:2018sht, Brandenberger:2003vk} for a review). This leaves us finally with a more compact format
   \begin{eqnarray}\label{dc1}
   \delta ^{\prime \prime}  + \mathcal{F}\mathcal{H} \delta ^{\prime}- \frac{3}{2}\mathcal{H}^2 \mathcal{G} \delta = 0,
   \end{eqnarray}
where we have defined the effective Hubble friction term $\mathcal{F}$ and the effective gravitational coupling $\mathcal{G}$, respectively:
\begin{eqnarray}\label{deltaprime}
   \mathcal{F} \equiv \frac{1}{\bar{\xi}}\biggl\{\biggl[4 - \frac{3}{\bar{\xi}}\left(1 + \frac{f_{\bar{T}\bar{T}}\bar{\rho}}{\kappa ^{2} + f_{\bar{T}}}\right)\biggr]\bar{\xi} + \frac{3}{\kappa ^{2} + f_{\bar{T}}}\biggl[f_{\bar{T}\bar{T}} - \frac{1}{\bar{\xi}}\left(f_{\bar{T}\bar{T}\bar{T}}\bar{\rho} + \frac{5}{2}f_{\bar{T}\bar{T}}\right)\biggr]\bar{\rho} + \frac{\bar{\xi}^{\prime}}{\mathcal{H}} - \frac{k ^{2}f_{\bar{T}}}{2\mathcal{H}(\kappa ^{2} + f_{\bar{T}})}\biggr\};
\end{eqnarray}
and
\begin{eqnarray}\label{delta}
    &&\mathcal{G} \equiv -\frac{2}{3\mathcal{H}^2\bar{\xi}} \biggl\{\frac{3}{\kappa ^{2} + f_{\bar{T}}}\biggl[f_{\bar{T}\bar{T}} - \frac{1}{\bar{\xi}}\biggl(f_{\bar{T}\bar{T}\bar{T}}\bar{\rho} + \frac{5}{2}f_{\bar{T}\bar{T}}\biggr)\biggr]\biggl[\mathcal{H}^{\prime}\bar{\rho} +\frac{3\mathcal{H}^2\bar{\rho}}{\bar{\xi}}\biggl(1 - \frac{f_{\bar{T}\bar{T}}}{\kappa ^{2} + f_{\bar{T}}}\biggr)\biggr] + \frac{3\mathcal{H}^{2}\bar{\rho}}{\kappa ^{2} + f_{\bar{T}}}\biggl[4 - \nonumber \\
    && - \frac{3}{\bar{\xi}}\left(1 + \frac{f_{\bar{T}\bar{T}}\bar{\rho}}{\kappa ^{2} + f_{\bar{T}}}\right)\biggr]\biggl[f_{\bar{T}\bar{T}} - \frac{1}{\bar{\xi}}\biggl(f_{\bar{T}\bar{T}\bar{T}}\bar{\rho} + \frac{5}{2}f_{\bar{T}\bar{T}}\biggr)\biggr] + \frac{3\mathcal{H}\bar{\rho}}{\kappa ^{2} + f_{T}}\biggl[-\frac{3\mathcal{H}\bar{\rho}}{\bar{\xi}}f_{\bar{T}\bar{T}\bar{T}} + \frac{\bar{\xi}^{\prime}}{\bar{\xi}^{2}}\biggl(\frac{5}{2}f_{\bar{T}\bar{T}} + f_{\bar{T}\bar{T}\bar{T}}\bar{\rho}\biggr) + \nonumber \\ 
    && + \frac{3\mathcal{H}\bar{\rho}}{\bar{\xi}^2}\biggl(f_{\bar{T}\bar{T}\bar{T}\bar{T}}\bar{\rho}  + \frac{7}{2}f_{\bar{T}\bar{T}\bar{T}}\biggr)\biggr] -  \frac{3k^{2}\mathcal{H}}{2(\kappa ^{2} + f_{\bar{T}})\bar{\xi}}\left(f_{\bar{T}\bar{T}}\bar{\rho} + f_{\bar{T}}\right) - \frac{3}{2}\mathcal{H}^2\left[1 + \frac{1}{\kappa ^{2}}\left(f_{\bar{T}\bar{T}}\bar{\rho} + \frac{3}{2}f_{\bar{T}}\right)\right]\biggr\}.
\end{eqnarray}
In \eqref{dc1}, by recovering GR with $f(R,T) = R$, then $\mathcal{F} = 1$ and $\mathcal{G} = 1$ as expected. Our task now is to assess how the chosen $f(R,T)$ models impact the evolution of the linear scalar perturbations.

\section{Numerical results and data analysis}

We proceed now solving Eq. \eqref{dc1} for competitive $f(R,T)$ models. We will test two different approaches i.e., the exponential one 
\begin{equation}\label{m1}
    f(R, T) = R + \alpha e^{\beta T},
\end{equation}
and the polynomial one
 \begin{equation}\label{m2}
     f(R, T) = R + \gamma _{n}T^{n}.
 \end{equation}
These models cover all existing proposed $f(R,T)$ models in literature. Hereafter, we will study them separately i.e., we either apply the exponential one or the polynomial one. We treat them as separate cases. Each model has the GR limit when either $\alpha$ or $\gamma_n$ vanish, respectively. In the exponential model, if $\alpha$ has a non-vanishing value and the trace dependence disappears with $\beta=0$, then the $\Lambda$CDM model is achieved. The later limit is also obtained in the polynomial model with $\gamma_n \neq 0$ and $n=0$. 

For each model there are two free parameters. One more than the flat $\Lambda$CDM model. Both parameters will be redefined as dimensionless quantities (denoted by an overbar) according to the definitions
\begin{equation}\label{parameters}
    \bar{\alpha} = \frac{\alpha }{\kappa ^{2}\rho _{0}}; \hspace{1cm} \bar{\beta} = \beta \rho _{0}; \hspace{1cm} \text{and} \hspace{1cm} \bar{\gamma}_{n} = \frac{\gamma _{n}\rho _{0}^{n - 1}}{\kappa ^{2}}.
\end{equation}

But what are the preferred free model parameters according to current cosmological data? This issue has been addressed recently in Ref. \cite{Jeakel:2023xlp}.  This reference has used the gas mass fraction in galaxy clusters, the background cosmic chronometers data and the age of the universe to test these two models. We have shown in the aforementioned reference these observations are enough to select viable cosmological scenario within $f(R,T)$ cosmologies. This analysis has shown that the late time dynamics can be fairly described by $f(R,T)$ gravity but only for a narrow range of the free model parameters. For both models the accepted parameter region is the one in which the trace dependence is quite weak (i.e., $\bar{\beta}$ and $n$ values close to zero), making these models very close to the $\Lambda$CDM cosmology. This is indeed a expected result. For every alternative cosmological scenario, by remaining close to the dynamics provided by the concordance flat $\Lambda$CDM model one can easily fit available data. However, in the case of modified gravity, no extra dark energy component is necessary. The "close to $\Lambda$ contribution" comes from the geometrical terms originated in the gravitational Lagrangian i.e., only the self-acceleration mode is able to explain the dark energy phenomena. We will explore latter $\bar{\beta}$ and $n$ values within the upper limit established by Ref. \cite{Jeakel:2023xlp}. More specifically, this means for the polynomial model $n\sim 0$ and $\bar{\gamma}_n \sim 1.4$. Remind that in the limit $n=0$ the condition $\bar{\gamma}_n=2 \Omega_{\Lambda}$ restores the $\Lambda$CDM model. The exponential model also remains very close to the $\lambda$ with $\bar{\beta} \sim 0$ and $\bar{\alpha} \sim 1.4$. However, in the exponential model, the statistical analysis performed with the gas mass fraction in galaxy clusters, the background cosmic chronometers data and the age of the universe data provided accepted $\bar{\alpha}$ and $\bar{\beta}$ values (at $2\sigma$ statistical confidence level) beyond the region close to the $\Lambda$CDM model. For example, combinations like $\{\bar{\alpha},\bar{\beta} \}=\{1.2,0.3\}, \{1.1,0.45\}$ and $\{1,0.6\}$ are allowed by available background observational data. We ask the reader to check Figs. $1$ and $2$ of Ref. \cite{Jeakel:2023xlp} for such results.

Given the complexity of expressions presented in \eqref{deltaprime} and \eqref{delta} the analytical treatment of \eqref{dc1} is not viable. We have then to solve this equation numerically for both models \eqref{m1}-\eqref{m2}. 

Our equation \eqref{dc1} for the evolution of $\delta$ is scale-dependent and valid in the linear regime. It is expected to break down for non-linear modes. We therefore adopt $k = 0.1hMpc^{-1}$ for the wavenumber appearing in \eqref{deltaprime} and \eqref{delta} in order to keep safely within the linear regime. This choice also maximizes the contribution of the $k^2$ dependent terms in \eqref{deltaprime} and \eqref{delta}. In all of our results the initial conditions are taken at the matter-radiation equality and they are set with the help of the CAMB code \cite{Lewis:1999bs} for the scale $k = 0.1hMpc^{-1}$. Specifically, we fix initial conditions for the density contrast of order $\delta(z_{eq}) \sim 10^{-4}$ where $z_{eq}=3400$ refers to the matter-radiation equality in the concordance $\Lambda$CDM cosmology. This assumption simplifies the analysis and is widely adopted in the literature. 

\begin{figure}[t!]
\center
\includegraphics[width=12cm]{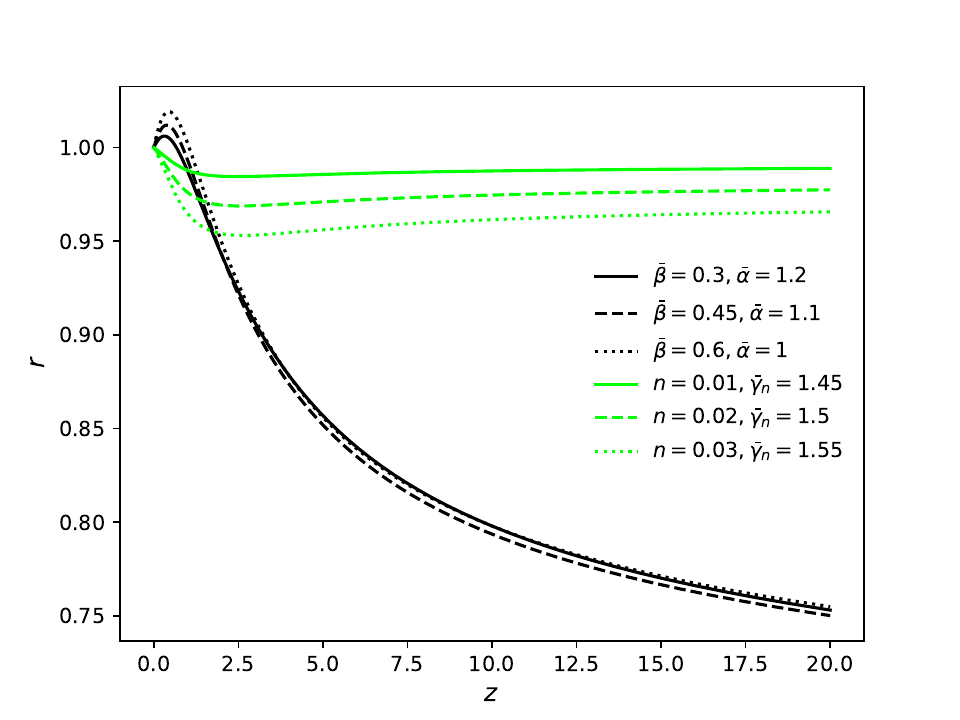}
\caption{Curves for the Alcock-Paczynski effect for the power law and exponential models for the region restricted by \cite{Jeakel:2023xlp}. The correction is of the order of 1\% when working with $\bar{\beta}(n) \sim 10^{-3}$ and $10 ^{-2}$ (see the corrected $f\sigma 8$ curves in Figure \ref{fsigma8}). }
\label{apeffect}
\end{figure}

After calculating the evolution of the matter density contrast $\delta$, we can compare the perturbative dynamics of $f(R,T)$ models with recent available data from LSS surveys. In this task, the relevant quantity is the combination $f \sigma_8$ where $f$ is the growth rate given by
\begin{equation}
  f\equiv \frac{d\ln{\delta}}{d\ln{a}}, 
\end{equation}
while $\sigma_8^2$ is the variance of the perturbed cosmic density field smoothed within a sphere of radius $8h^{-1}Mpc$ calculated in the linear regime as
\begin{equation}
    \sigma _{8}(z) = \frac{\delta (z)}{\delta (z = 0)}\sigma _{8} (z = 0). 
\end{equation}
As a function of the redshift this combination reads
\begin{equation}
    f\sigma_{8}(z) = -(1 + z)\frac{\sigma _{8}(z = 0)}{\delta (z = 0)}\frac{d}{dz}\delta (z).
\end{equation}
Since we are dealing with cosmologies that significantly deviates from the fiducial flat $\Lambda$CDM model, one has to correct the theoretical $f \sigma_8$ with the Alcock-Paczynski effect. Following for example, Ref. \cite{nesseris}, this correction is implemented by multiplying the above expression by the ratio
\begin{equation}
        r=\frac{H^{model} D^{model}_A}{H^{fid}D^{fid}_A},
\end{equation}
where $H$ is the Hubble expansion and $D_A$ the angular distance. As we show in Fig. \eqref{apeffect} the correction $r$ is not so relevant (less than $5\%$) for the power law model, but it is relevant in the exponential model.

One can also calculate the so called $S_{8}$ parameter defined according to
\begin{equation}
    S_{8} = \sigma _{8}(z)\left(\frac{\Omega _{m}}{0.3}\right)^{1/2}.
\end{equation}

Current attempts to measure this parameter have revealed an apparent cosmological tension. Inferred $S_8$ values from the Cosmic Microwave Background (CMB) data obtained by the PLANCK mission ($S_8=0.83\pm 0.013$) are in disagreement with estimations from galaxy surveys like eBOSS (Extended Baryon Oscillation Spectroscopic Survey) $S_8=0.728\pm 0.026$ \cite{eBOSS:2020yzd}, DES (Dark Energy Survey)  $S_8=0.783\pm 0.021$ \cite{DES:2017myr} or KiDS (Kilo-Degree Survey) $S_8=0.766\pm 0.020$ \cite{Heymans:2020gsg}. These surveys provide a lower value for $S_8$ with relevant statistical confidence level than the one inferred by CMB data. Since $S_8$ quantifies the amplitude of the variance of matter fluctuations (see \cite{Nunes:2021ipq} for a current discussion about this observable), there is the possibility to have the following interpretation: galactic surveys are indicating a suppressed clustering growth. This same trend has also been recently pointed out in Ref. \cite{Nguyen:2023fip} via the direct fitting of available $f \sigma_8$ data. These results will be used below to select $f(R,T)$ models that have such clustering growth suppression. When calculating the $S_8$ values for the $f(R,T)$ models will adopt the following strategy. With the new evolution for the density contrast given by \eqref{dc1} and obtain a modified value for $\sigma_8$ which directs affects the resulting $S_8$ value.

\begin{figure}[t!]
\center
\includegraphics[width=12cm]{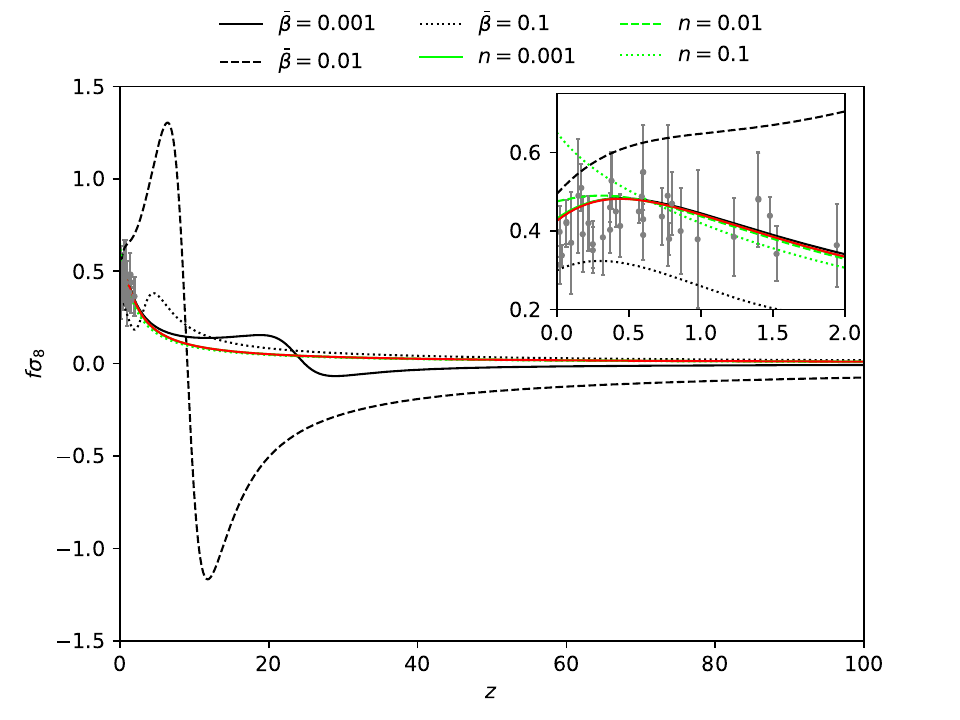}
\caption{Evolution of $f \sigma_8$ as a function of the redshift. The data points are taken from Refs. \cite{Benisty:2020kdt, Said:2020epb, Kazantzidis:2018rnb,Sakr:2023bms}. The black (green) lines refer to different $\bar{\beta} (n)$ values for the exponential (polynomial) model fixing $\bar{\alpha} (\bar{\gamma}_n)=1.4$. The inset amplifies the low redshift region.}
\label{fsigma8}
\end{figure}

We show in Fig. \eqref{fsigma8} the evolution of the $f\sigma_8$ combination as a function of the redshift in the studied $f(R,T)$ models. The red curve corresponds to the flat $\Lambda$CDM model with $\Omega_{\Lambda}=0.7$. The curves for the exponential $f(R,T)$ model are displayed in the black lines. The polynomial one in green. For both models we fix the parameters $\bar{\alpha}=\bar{\gamma}_n=1.4$ and vary either $\bar{\beta}$ (the exponential model) or $n$ (the polynomial model). The values used for $\bar{\beta}$ and $n$ are shown in the figure. Available observational data  shown in gray is obtained from Refs. \cite{Benisty:2020kdt, Said:2020epb, Kazantzidis:2018rnb,Sakr:2023bms}. The inset of this figure amplifies the low redshift region. The qualitative behavior seen here has a general trend. The polynomial model (green lines) remains quite similar to the $\Lambda$CDM model in the past but it has an amplified growth at low redshift. This feature is better seen in the inset of Fig. \eqref{fsigma8}. The exponential models present the oscillatory feature at redshifts of order $10$ to $20$, depending on the $\bar{\beta}$ values. The solid black line, corresponding to $\bar{\beta}=0.001$, is identical to the red one ($\Lambda$CDM) at low redshfits. However, even with such small value $\bar{\beta}=0.001$ there is an oscillatory feature at $z \sim 20$.

\begin{figure}[h!]
\center
\includegraphics[width=12cm]{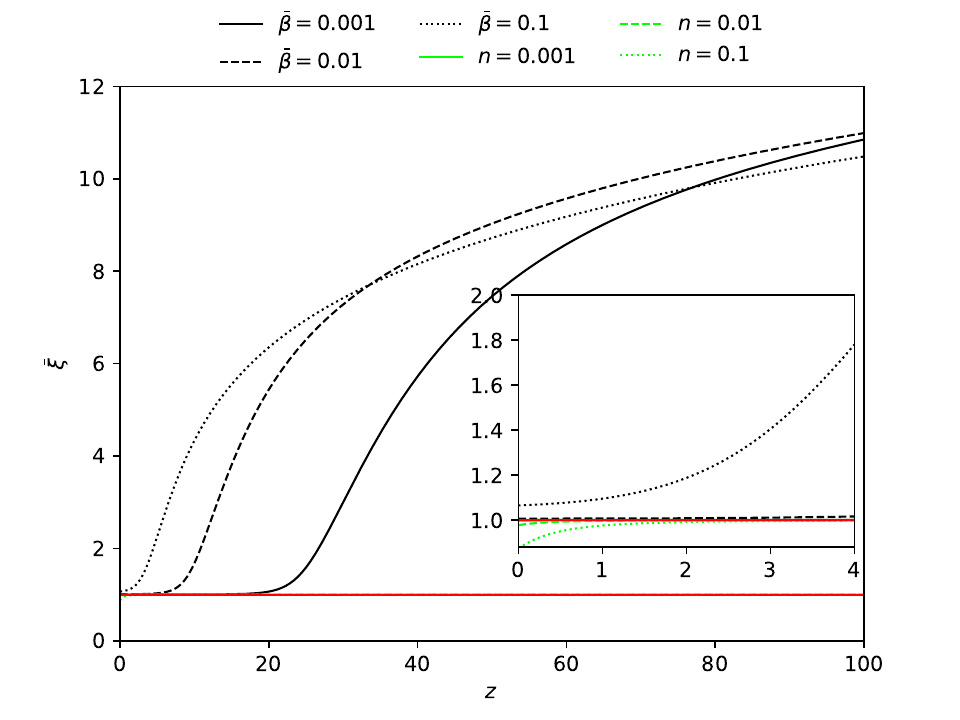}
\caption{Evolution of the parameter $\bar{\xi}$, equation \eqref{csi}, as a function of the redshift. Black lines represent the exponential model where we have fixed $\bar{\alpha}=1.4$. For the polynomial models (green lines) we have fixed $\bar{\gamma}_n=1.4$. The red line is flat $\Lambda$CDM limit with $\Omega_{\Lambda}=0.7$ ($\bar{\beta}$ ($n$) = 0). }
\label{figurecsi}
\end{figure}

\begin{figure}[t!]
\center
\includegraphics[width=12cm]{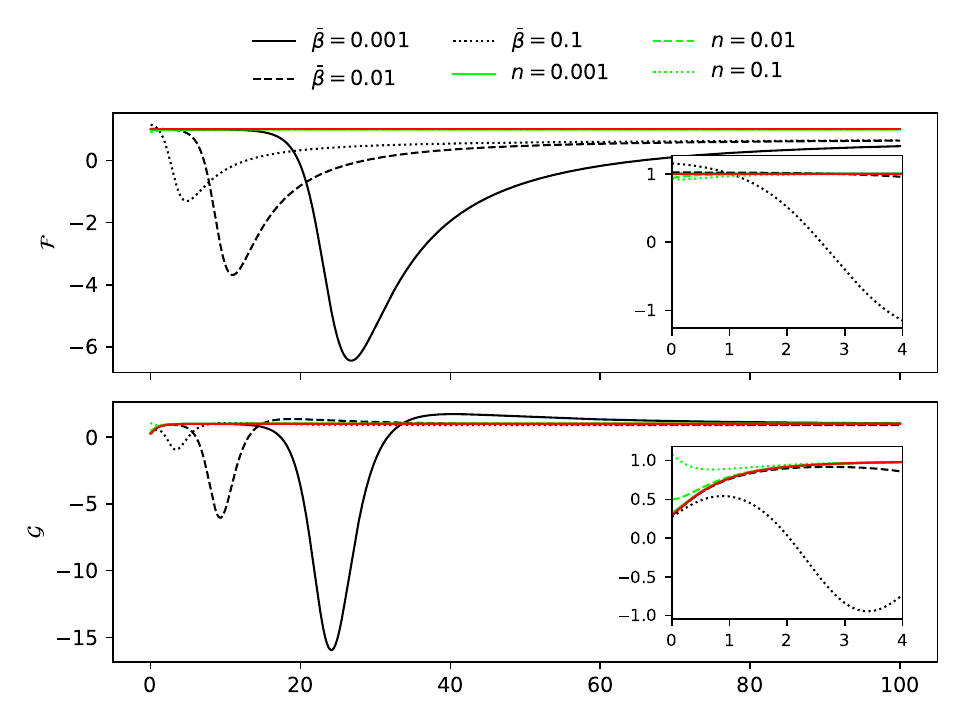}
\caption{Evolution of the term effective Hubble friction term $\mathcal{F}$, equation \eqref{deltaprime}, and the funtion $\mathcal{G}$, equation \eqref{delta}, as a function of the redshift. }
\label{figurefg}
\end{figure}

For a better visualization of how the different modified gravity model parameters influence the evolution of the density contrast given by Eq. \eqref{dc1} let us plot the functions $\bar{\xi}$ \eqref{csi}, $\mathcal{F}$ \eqref{deltaprime} and $\mathcal{G}$ \eqref{delta}. They are displayed respectively in Fig. \eqref{figurecsi}, in the top panel of \eqref{figurefg} and in the botton panel of \eqref{figurefg}.

The behavior of $\bar{\xi}$ seen in \eqref{figurecsi} influences the conservation law \eqref{conserxi}. For the different $n$ values used in the polynomial model, the resulting curves are almost indistinguishable from GR. Again, the adopted values for the parameter $n$ are based on the results presented in Ref. \cite{Jeakel:2023xlp} i.e., only $n \lesssim 0.1$ are observationally allowed. Conversely, the exponential model has a quite different evolution. In the exponential model, by keeping $\bar{\alpha}=1.4$, only $\bar{\beta}\lesssim 0.1$ are allowed from the background analysis. With such range of values for $\bar{\alpha}$ and $\bar{\beta}$, the resulting behavior of $\bar{\xi}$ is shown in the black lines of \eqref{figurecsi}. The quantity $\bar{\xi}$ increases for high redshifts. This makes the non-conservation aspect more relevant during the early stage of the universe. Apart from the numerical visualization provided by this figure, it is also possible to have an analytical hint about this behavior. By substituting the $f(R,T)$ models \eqref{m1} and \eqref{m2} into \eqref{csi} one finds that in the high density (past) limit the $\bar{\xi}$ quantity tends to a constant value in the polynomial model. This constant value differs from the unity proportionally to $n$. However, for the exponential model, $\bar{\xi}$ is proportional to the matter density parameter $\Omega$ which is a monotonically increasing function towards high redshifts. This explains the behavior observed in Fig. \eqref{figurecsi}.

The difference between the polynomial and the exponential models is also clear when analysing the curves for $\mathcal{F}$ and $\mathcal{G}$ in Fig. \eqref{figurefg}. The curves for the exponential models bounces at high redshift values. The smaller the $\bar{\beta}$ value, later this bounce happens and the amplitude of oscillation becomes smaller. We have checked this bounce is related to the change in the sign of the derivative of the $\bar{\xi}$ function as shown in Fig. \eqref{figurecsi}. A visual inspection of both figures reveals this feature. 

\begin{figure}[t!]
\center
\includegraphics[width=12cm]{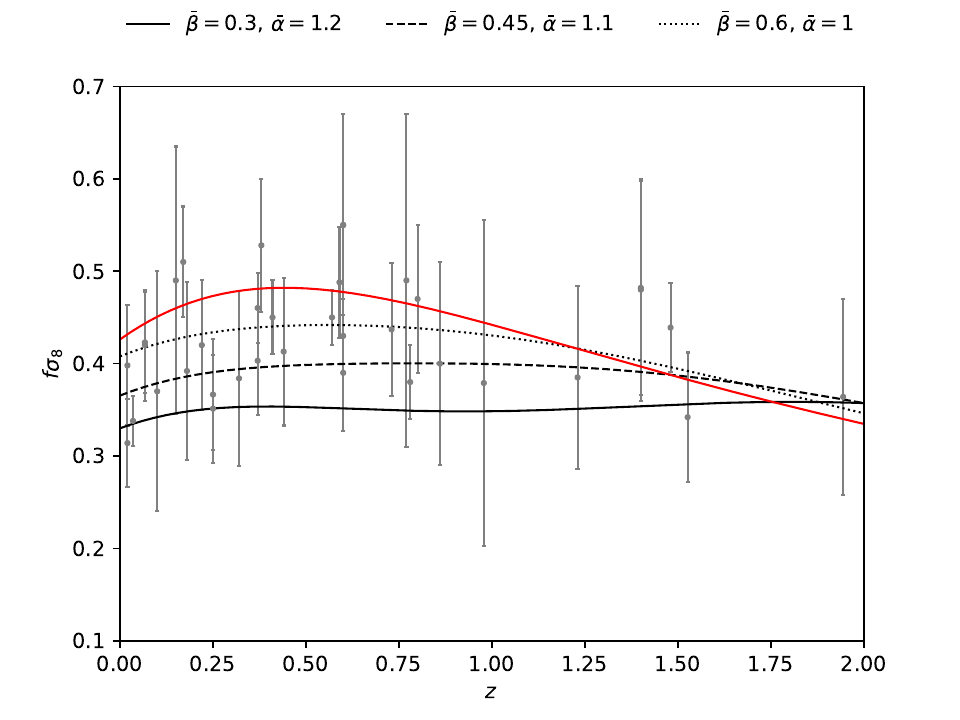}
\caption{Evolution of $f \sigma_8$ as a function of the redshift. The data points are taken from Refs. \cite{Benisty:2020kdt, Said:2020epb, Kazantzidis:2018rnb,Sakr:2023bms}. The black lines refer to different combinations of $\bar{\alpha}$ and $\bar{\beta} $ parameters for the exponential model.}
\label{fs8final}
\end{figure}

We have so far investigated the free model parameters in which the studied $f(R,T)$ models are very close to the $\Lambda$CDM at background level. However, concerning the exponential model, even such small $\bar{\beta}$ values are able to leave pathological imprints like oscillations on the $f\sigma_8$ observable. We remind now that according to Ref. \cite{Jeakel:2023xlp} the observationally allowed free space parameters in the exponential model can be extended to regions beyond the $\Lambda$CDM limit. As mentioned before, the combinations like $\{\bar{\alpha},\bar{\beta} \}=\{1.2,0.3\}, \{1.1,0.45\}$ and $\{1,0.6\}$ are in agreement with background cosmological data. These $\{\bar{\alpha},\bar{\beta} \}$ combinations and the $\Lambda$CDM limit characterized by $\{\bar{\alpha},\bar{\beta} \}=\{1.4,0\}.$ are in agreement with background cosmological data within $3\sigma$ of confidence level as shown in \cite{Jeakel:2023xlp}. Indeed, in terms of statistical tests that take into account the number of model free parameters, the  exponential $f(R,T)$ model with $3$ free parameters, ($H_0, \bar{\alpha}, \bar{\beta}$), will be severely penalized against the flat $\Lambda$CDM with only 2 free parameters ($H_0, \Omega_{m0}$). We test therefore the perturbative dynamics with this set of free model parameter values in the exponential model. We show in Fig. \eqref{fs8final} the $f\sigma_8$ curves for these parameter values (black lines) in comparison to the flat $\Lambda$CDM model with $\Omega_{\Lambda}=0.7$ (the red curve) obtained in the limit $\bar{\beta}=0$ and $\bar{\alpha}=1.4$. Both black curves have a slightly larger $f \sigma_8$ amplitude at high redshifts but they transit to a suppressed amplitude at small redshifts, in the region where observational data is available.

Apart from the analysis using the $f \sigma_8$ observable, let us also use equations for the gravitational potential \eqref{potential}, derived in the previous section, in order to compute the Integrated Sachs-Wolfe (ISW) effect of both $f(R,T)$ models and compare it with 
the prediction of a flat $\Lambda$CDM model with $\Omega_{m0}=0.3$. The $H_0$ parameter plays no role in this analysis. We calculate the relative amplifications ($Q$) of the ISW effect as

\begin{equation}
    Q=\frac{\left(\frac{\Delta T}{T}\right)^{f(R,T)}_{ISW}}{\left(\frac{\Delta T}{T}\right)^{\Lambda CDM}_{ISW}}-1.
\end{equation}

If $Q > 0$ $(< 0)$ the $f(R,T)$ models produce more (less) temperature variation to the CMB
photons via the ISW effect than the adopted $\Lambda$CDM model. Such an analysis of the ISW effect has been proposed in \cite{Dent:2008ek} and used in \cite{Velten:2011bg,Velten:2012uv}. Due to our previous results, we present only the results for the exponential model. With the same parameter values as used in \eqref{fs8final} the $Q$ values for $\{\bar{\alpha},\bar{\beta} \}=\{1.2,0.3\}, \{1.1,0.45\}$ and $\{1,0.6\}$ are, respectively, $32.9 \%, 29.9 \%,$ and $ 24.0\%$. This means that all tested values enhance the ISW signal in comparison to the $\Lambda$CDM model. Such level of enhancement can not be probed by current cross-correlation studies aiming to detect the ISW effect (see, for example, \cite{Bahr-Kalus:2022yny}).

\section{Conclusions}

We have studied $f(R,T)$ modified gravity models, being $T$ the trace of the energy-momentum tensor. We have focused on the analysis of cosmological linear scalar perturbations. The equation for the evolution of the matter density contrast $\delta$ has been presented in the compact form \eqref{dc1}. We remark our equation is different from the one previously found in Ref. \cite{Alvarenga:2013syu} even adopting the same approximation, the quasi-static one. We have then studied the perturbative behavior of the exponential $f(R,T)$ model \eqref{m1} and the polynomial one \eqref{m2}. 

 In order to assess the viability of such models we have used the $f \sigma_8$ observable. The polynomial model presents an excess of clustering amplitude at low redshifts leading to high $f\sigma_8$ values (green lines in Fig. \eqref{fsigma8}). This behavior is challenged by the recent findings of Ref. \cite{Nguyen:2023fip} which indicates that a suppressed clustering is preferred by data. Therefore, we can conclude the polynomial model is not a competitive $f(R,T)$ model in light of $f \sigma_8$ data.

 Concerning the exponential model, our first analysis focused on the free parameter space near the flat $\Lambda$CDM model i.e., the limit $\beta \sim 0$ and $\bar{\alpha}=1.4$. The latter value corresponds to $\Omega_{\Lambda}=0.7$ when $\bar{\beta}$ vanishes. With the parameter values near the $\Lambda$CDM limit we found oscillations in the $f \sigma_8$ evolution (black lines of Fig. \eqref{fsigma8}). This pathological behavior also compromises the viability of such parameter values. We have then moved to another region of the free model parameters, by testing the set $\{\bar{\alpha},\bar{\beta} \}=\{1.2,0.3\}, \{1.1,0.45\}$ and $\{1,0.6\}$, also allowed at background level as shown in Ref. \cite{Jeakel:2023xlp}. The corresponding $f\sigma_8$ curves for this set of parameter values, as shown in Fig. \eqref{fs8final}, are in good agreement with observational data. The calculated $S_8$ values are, respectively, $0.81, 0.75$ and $0.75$, alleviating the $S_8$ tension according to the observational values quoted below (41).

 In conclusion, we have extended in this work the analysis performed in Ref. \cite{Jeakel:2023xlp} to the scalar linear perturbations. Now, we are finally able to exclude the polynomial $f(R,T)$ model as a competitive candidate to explain the late time accelerated cosmological expansion. Only with an almost vanishing $T$ dependence given by $n<0.1$ the models would fit both background and linear clustering data. On the other hand, the exponential model has passed both tests, but its allowed free model parameter space has been severely constrained. Even so, in the latter scenarios there is a mild enhancement in the ISW effect signal that can be constrained with future surveys. In this case, the exponential model can be further refined with Markov Chain Monte Carlo analysis with all available data in order to explore in great detail its phase space parameters in a more quantitative way. This can be done in a future work.

\section*{Acknowledgments} Partial support from Brazilian funding agencies PROPPi/UFOP, CAPES, CNPq, FAPEMIG and FAPES is acknowledged. We thank Sunny Vagnozzi and Radouane Gannouji for useful comments.


\begin{thebibliography}{99}

\bibitem{Velten:2017hhf}
H.~Velten and T.~R.~P.~Caram\^es,
Phys. Rev. D \textbf{95} (2017) no.12, 123536, 
\href{https://journals.aps.org/prd/abstract/10.1103/PhysRevD.95.123536}{doi:10.1103/PhysRevD.95.123536},   
 \href{https://arxiv.org/abs/1702.07710}{[arXiv:1702.07710 [gr-qc]]}.

\bibitem{Moraes:2019hgx}
P.~H.~R.~S.~Moraes, P.~K.~Sahoo and S.~K.~J.~Pacif,
Gen. Rel. Grav. \textbf{52} (2020) no.4, 32, 
\href{https://link.springer.com/article/10.1007/s10714-020-02681-3}{doi:10.1007/s10714-020-02681-3},  
\href{https://arxiv.org/abs/1905.00417}{[arXiv:1905.00417 [gr-qc]]}.

\bibitem{Jeakel:2023xlp}
A.~P.~Jeakel, J.~Pinheiro da Silva and H.~Velten,
Phys. Dark Univ. \textbf{43}, 101401 (2024), 
\href{https://www.sciencedirect.com/science/article/pii/S2212686423002340}{doi:10.1016/j.dark.2023.101401}, 
\href{https://arxiv.org/abs/2303.15208}{[arXiv:2303.15208v3 [astro-ph.CO]]}


\bibitem{Myrzakulov:2023lcl}
N.~Myrzakulov, M.~Koussour, A.~H.~A.~Alfedeel and E.~I.~Hassan,
Chin. Phys. C \textbf{47}, no.11, 115107 (2023), \href{https://iopscience.iop.org/article/10.1088/1674-1137/acf2fa}{doi:10.1088/1674-1137/acf2fa}, 
\href{https://arxiv.org/abs/2308.09913}{[arXiv:2308.09913 [astro-ph.CO]]}.

\bibitem{Iosifidis:2021kqo}
D.~Iosifidis, N.~Myrzakulov and R.~Myrzakulov,
Universe \textbf{7} (2021) no.8, 262
\href{https://www.mdpi.com/2218-1997/7/8/262}{doi:10.3390/universe7080262}, 
\href{https://arxiv.org/abs/2106.05083}{[arXiv:2106.05083 [gr-qc]]}.

\bibitem{Fortunato:2023ypc}
J.~A.~S.~Fortunato, P.~H.~R.~S.~Moraes, J.~G.~d.~J\'unior and E.~Brito,
Eur. Phys. J. C \textbf{84}, no.2, 198 (2024), 
\href{https://link.springer.com/article/10.1140/epjc/s10052-024-12544-9}{doi:10.1140/epjc/s10052-024-12544-9}, 
\href{https://arxiv.org/abs/2305.01325}{[arXiv:2305.01325 [gr-qc]]}.

\bibitem{Mishra:2024uwq}
R.~K.~Mishra and N.~Jain,
Int. J. Theor. Phys. \textbf{63}, no.1, 29 (2024), 
\href{https://link.springer.com/article/10.1007/s10773-024-05553-7}{doi:10.1007/s10773-024-05553-7}

\bibitem{Bhattacharjee:2019oim}
S.~Bhattacharjee and P.~K.~Sahoo,
Grav. Cosmol. \textbf{26}, no.3, 281-284 (2020), 
\href{https://link.springer.com/article/10.1134/S0202289320030032}{doi:10.1134/S0202289320030032}, 
\href{https://arxiv.org/abs/1908.06759}{[arXiv:1908.06759 [gr-qc]]}.

\bibitem{Bouali:2023fid}
A.~Bouali, H.~Chaudhary, T.~Harko, F.~S.~N.~Lobo, T.~Ouali and M.~A.~S.~Pinto,
Mon. Not. Roy. Astron. Soc. \textbf{526}, 4192-4208 (2023), 
\href{https://academic.oup.com/mnras/article/526/3/4192/7313617}{doi:10.1093/mnras/stad2998}, 
\href{https://arxiv.org/abs/2309.15497}{[arXiv:2309.15497 [gr-qc]]}.

\bibitem{Bertini:2023pmp}
N.~R.~Bertini and H.~Velten,
Phys. Rev. D \textbf{107} (2023) no.12, 124005, 
\href{https://journals.aps.org/prd/abstract/10.1103/PhysRevD.107.124005}{doi:10.1103/PhysRevD.107.124005}, 
\href{https://arxiv.org/abs/2303.09699}{[arXiv:2303.09699 [gr-qc]]}.


\bibitem{Fisher:2019ekh}
S.~B.~Fisher and E.~D.~Carlson,
Phys. Rev. D \textbf{100}, no.6, 064059 (2019), 
\href{https://journals.aps.org/prd/abstract/10.1103/PhysRevD.100.064059}{doi:10.1103/PhysRevD.100.064059}, 
\href{https://arxiv.org/abs/1908.05306 }{[arXiv:1908.05306 [gr-qc]]}.

\bibitem{PhysRevD.101.108501}Harko, T. \& Moraes, P. Comment on “Reexamining f(R,T) gravity”. {\em Phys. Rev. D}. \textbf{101}, 108501 (2020,5), \href{https://journals.aps.org/prd/abstract/10.1103/PhysRevD.101.108501}{doi:10.1103/PhysRevD.101.108501}.

\bibitem{Fisher:2020zwx}
S.~B.~Fisher and E.~D.~Carlson,
Phys. Rev. D \textbf{101}, no.10, 108502 (2020), 
\href{https://journals.aps.org/prd/abstract/10.1103/PhysRevD.101.108502}{doi:10.1103/PhysRevD.101.108502}, 
\href{https://arxiv.org/abs/2004.02934}{[arXiv:2004.02934 [gr-qc]]}.

\bibitem{Shabani:2014xvi}
H.~Shabani and M.~Farhoudi,
Phys. Rev. D \textbf{90}, no.4, 044031 (2014), 
\href{https://journals.aps.org/prd/abstract/10.1103/PhysRevD.90.044031}{doi:10.1103/PhysRevD.90.044031}, 
\href{https://arxiv.org/abs/1407.6187}{[arXiv:1407.6187 [gr-qc]]}.

\bibitem{Alvarenga:2013syu}
F.~G.~Alvarenga, A.~de la Cruz-Dombriz, M.~J.~S.~Houndjo, M.~E.~Rodrigues and D.~S\'aez-G\'omez,
Phys. Rev. D \textbf{87} (2013) no.10, 103526
[erratum: Phys. Rev. D \textbf{87} (2013) no.12, 129905], 
\href{https://journals.aps.org/prd/abstract/10.1103/PhysRevD.87.103526}{doi:10.1103/PhysRevD.87.103526}, 
\href{https://arxiv.org/abs/1302.1866}{[arXiv:1302.1866 [gr-qc]]}.


\bibitem{Harko:2011kv}
T.~Harko, F.~S.~N.~Lobo, S.~Nojiri and S.~D.~Odintsov,
Phys. Rev. D \textbf{84} (2011), 024020, 
\href{https://journals.aps.org/prd/abstract/10.1103/PhysRevD.84.024020}{doi:10.1103/PhysRevD.84.024020}, 
\href{https://arxiv.org/abs/1104.2669}{[arXiv:1104.2669 [gr-qc]]}.

\bibitem{Barrientos:2018cnx}
E.~Barrientos, F.~S.~N.~Lobo, S.~Mendoza, G.~J.~Olmo and D.~Rubiera-Garcia,
Phys. Rev. D \textbf{97} (2018) no.10, 104041, 
\href{https://journals.aps.org/prd/abstract/10.1103/PhysRevD.97.104041}{doi:10.1103/PhysRevD.97.104041}, 
\href{https://arxiv.org/abs/1803.05525}{[arXiv:1803.05525 [gr-qc]]}.

\bibitem{Sotiriou:2008it}
T.~P.~Sotiriou and V.~Faraoni,
Class. Quant. Grav. \textbf{25}, 205002 (2008), 
\href{https://iopscience.iop.org/article/10.1088/0264-9381/25/20/205002}{doi:10.1088/0264-9381/25/20/205002}, 
\href{https://arxiv.org/abs/0805.1249}{[arXiv:0805.1249 [gr-qc]]}.

\bibitem{Bertolami:2008ab}
O.~Bertolami, F.~S.~N.~Lobo and J.~Paramos,
Phys. Rev. D \textbf{78}, 064036 (2008), 
\href{https://journals.aps.org/prd/abstract/10.1103/PhysRevD.78.064036}{doi:10.1103/PhysRevD.78.064036}, 
\href{https://arxiv.org/abs/0806.4434}{[arXiv:0806.4434 [gr-qc]]}.


\bibitem{Faraoni:2009rk}
V.~Faraoni,
Phys. Rev. D \textbf{80}, 124040 (2009), 
\href{https://journals.aps.org/prd/abstract/10.1103/PhysRevD.80.124040}{doi:10.1103/PhysRevD.80.124040}, 
\href{https://arxiv.org/abs/0912.1249}{[arXiv:0912.1249 [astro-ph.GA]]}.

\bibitem{Mukhanov:1988jd}
V.~F.~Mukhanov,
Sov. Phys. JETP \textbf{67} (1988), 1297-1302.

\bibitem{Brandenberger:2003vk}
R.~H.~Brandenberger,
Lect. Notes Phys. \textbf{646} (2004), 127-167, 
\href{https://link.springer.com/chapter/10.1007/978-3-540-40918-2_5}{doi:10.1007/978-3-540-40918-2\_5},  
\href{https://arxiv.org/abs/hep-th/0306071}{[arXiv:hep-th/0306071 [hep-th]]}.

\bibitem{Maggiore:2018sht}
M.~Maggiore,
Oxford University Press, 2018,
ISBN 978-0-19-857089-9

\bibitem{Lewis:1999bs}
A.~Lewis, A.~Challinor and A.~Lasenby,
Astrophys. J. \textbf{538}, 473-476 (2000), 
\href{https://iopscience.iop.org/article/10.1086/309179}{doi:10.1086/309179}, 
\href{https://arxiv.org/abs/astro-ph/9911177}{[arXiv:astro-ph/9911177 [astro-ph]]}.

\bibitem{nesseris}
S.~Nesseris, G.~Pantazis and L.~Perivolaropoulos,
Phys. Rev. D \textbf{96} (2017) no.2, 023542, \href{https://journals.aps.org/prd/abstract/10.1103/PhysRevD.96.023542}{doi:10.1103/PhysRevD.96.023542}, \href{https://arxiv.org/abs/1703.10538}{[arXiv:1703.10538 [astro-ph.CO]]}. 

\bibitem{eBOSS:2020yzd}
S.~Alam \textit{et al.} [eBOSS],
Phys. Rev. D \textbf{103} (2021) no.8, 083533, 
\href{https://journals.aps.org/prd/abstract/10.1103/PhysRevD.103.083533}{doi:10.1103/PhysRevD.103.083533}, 
\href{https://arxiv.org/abs/2007.08991}{[arXiv:2007.08991 [astro-ph.CO]]}.

\bibitem{DES:2017myr}
T.~M.~C.~Abbott \textit{et al.} [DES],
Phys. Rev. D \textbf{98} (2018) no.4, 043526, 
\href{https://journals.aps.org/prd/abstract/10.1103/PhysRevD.98.043526}{doi:10.1103/PhysRevD.98.043526}, 
\href{https://arxiv.org/abs/1708.01530}{[arXiv:1708.01530 [astro-ph.CO]]}.

\bibitem{Heymans:2020gsg}
C.~Heymans, T.~Tr\"oster, M.~Asgari, C.~Blake, H.~Hildebrandt, B.~Joachimi, K.~Kuijken, C.~A.~Lin, A.~G.~S\'anchez and J.~L.~van den Busch, \textit{et al.}
Astron. Astrophys. \textbf{646} (2021), A140, 
\href{https://www.aanda.org/articles/aa/full_html/2021/02/aa39063-20/aa39063-20.html}{doi:10.1051/0004-6361/202039063}, 
\href{https://arxiv.org/abs/2007.15632}{[arXiv:2007.15632 [astro-ph.CO]]}.

\bibitem{Nunes:2021ipq}
R.~C.~Nunes and S.~Vagnozzi,
Mon. Not. Roy. Astron. Soc. \textbf{505} (2021) no.4, 5427-5437, \href{https://arxiv.org/abs/2106.01208}{
doi:10.1093/mnras/stab1613}
{[arXiv:2106.01208 [astro-ph.CO]]}

\bibitem{Nguyen:2023fip}
N.~M.~Nguyen, D.~Huterer and Y.~Wen,
Phys. Rev. Lett. \textbf{131} (2023) no.11, 111001, 
\href{https://arxiv.org/abs/2302.01331}{doi:10.1103/PhysRevLett.131.111001}, 
\href{https://journals.aps.org/prl/abstract/10.1103/PhysRevLett.131.111001}{[arXiv:2302.01331 [astro-ph.CO]]}.


\bibitem{Benisty:2020kdt}
D.~Benisty,
Phys. Dark Univ. \textbf{31} (2021), 100766, 
\href{https://www.sciencedirect.com/science/article/pii/S2212686420304799?via%3Dihub}{doi:10.1016/j.dark.2020.100766}, 
\href{https://arxiv.org/abs/2005.03751}{[arXiv:2005.03751 [astro-ph.CO]]}.

\bibitem{Said:2020epb}
K.~Said, M.~Colless, C.~Magoulas, J.~R.~Lucey and M.~J.~Hudson,
Mon. Not. Roy. Astron. Soc. \textbf{497} (2020) no.1, 1275-1293, 
\href{https://academic.oup.com/mnras/article/497/1/1275/5870121}{doi:10.1093/mnras/staa2032}, 
\href{https://arxiv.org/abs/2007.04993}{[arXiv:2007.04993 [astro-ph.CO]]}.


\bibitem{Kazantzidis:2018rnb}
L.~Kazantzidis and L.~Perivolaropoulos,
Phys. Rev. D \textbf{97} (2018) no.10, 103503, 
\href{https://journals.aps.org/prd/abstract/10.1103/PhysRevD.97.103503}{doi:10.1103/PhysRevD.97.103503}, 
\href{https://arxiv.org/abs/1803.01337}{[arXiv:1803.01337 [astro-ph.CO]]}.

\bibitem{Sakr:2023bms}
Z.~Sakr,
Universe \textbf{9} (2023) no.8, 366, 
\href{https://www.mdpi.com/2218-1997/9/8/366}{doi:10.3390/universe9080366}, 
\href{https://arxiv.org/abs/2305.02863}{[arXiv:2305.02863 [astro-ph.CO]]}.


\bibitem{Dent:2008ek}
J.~B.~Dent, S.~Dutta and T.~J.~Weiler,
Phys. Rev. D \textbf{79} (2009), 023502,
\href{https://journals.aps.org/prd/abstract/10.1103/PhysRevD.79.023502}{doi:10.1103/PhysRevD.79.023502}, 
\href{https://arxiv.org/abs/0806.3760}{[arXiv:0806.3760 [astro-ph]]}.

\bibitem{Velten:2011bg}
H.~Velten and D.~J.~Schwarz,
JCAP \textbf{09} (2011), 016, 
\href{https://iopscience.iop.org/article/10.1088/1475-7516/2011/09/016}{doi:10.1088/1475-7516/2011/09/016}, 
\href{https://arxiv.org/abs/1107.1143}{[arXiv:1107.1143 [astro-ph.CO]]}.

\bibitem{Velten:2012uv}
H.~Velten and D.~Schwarz,
Phys. Rev. D \textbf{86} (2012), 083501, 
\href{https://journals.aps.org/prd/abstract/10.1103/PhysRevD.86.083501}{doi:10.1103/PhysRevD.86.083501}, 
\href{https://arxiv.org/abs/1206.0986}{[arXiv:1206.0986 [astro-ph.CO]]}.


\bibitem{Bahr-Kalus:2022yny}
B.~Bahr-Kalus, D.~Parkinson, J.~Asorey, S.~Camera, C.~Hale and F.~Qin,
Mon. Not. Roy. Astron. Soc. \textbf{517} (2022) no.3, 3785-3803, 
\href{https://academic.oup.com/mnras/article/517/3/3785/6652116}{doi:10.1093/mnras/stac2040}, 
\href{https://arxiv.org/abs/2204.13436}{[arXiv:2204.13436 [astro-ph.CO]]}
.
\end{thebibliography}
\end{document}